\documentclass[aps,twocolumn,showpacs,preprintnumbers,eqsecnum,amsmath,amssymb,nofootinbib]{revtex4}
\usepackage{graphics}
\usepackage{dcolumn}
\usepackage{bm}
\usepackage{epsfig}

\begin{document} 

\title{\boldmath Nonintegrable threshold singularities of two-point functions in perturbation theory}
\author{Dmitri Melikhov$^{a,b,c}$ and Silvano Simula$^d$}
\affiliation{ 
$^a$D.~V.~Skobeltsyn Institute of Nuclear Physics, M.~V.~Lomonosov Moscow State University, 119991, Moscow, Russia\\
$^b$Joint Institute for Nuclear Research, 141980 Dubna, Russia\\
$^c$Faculty of Physics, University of Vienna, Boltzmanngasse 5, A-1090 Vienna, Austria\\ 
$^d$Istituto Nazionale di Fisica Nucleare, Sezione di Roma Tre, Via della Vasca Navale 84, I-00146 Roma, Italy}
\begin{abstract}
In perturbation theory, the spectral densities of two-point functions develop nonintegrable threshold singularities
at higher orders. In QCD, such singularities emerge when calculating the diagrams in terms of the pole quark mass,
and they become stronger when one rearranges the perturbative expansion in terms of the running quark mass.
In this paper we discuss the proper way to handle such singularities.
\end{abstract}


\maketitle

\section{Introduction}
The correlation function of two currents defined as
\begin{eqnarray}
\label{Pi}
\Pi(q)=i \int dx e^{i q x}\langle T\{\bar \Psi(x){\cal O}_1\psi(x), \bar \psi(0){\cal O}_2\Psi(0)\}\rangle,
\nonumber\\
\end{eqnarray}
where $\psi$ and $\Psi$ denote fermion fields (which may be identical) and ${\cal O}_{1,2}$ are Dirac matrices,
is one of the basic objects in quantum field theory. For instance, in QCD, two-point functions with an appropriate choice of
the Dirac matrices provide the basis for the extraction of masses and couplings of mesons with the corresponding quantum
numbers within the method of QCD sum rules \cite{nosvvz,svz}. In general, the two point function contains a number of
independent Lorentz structures $L_i(q)$ and the corresponding invariant amplitudes $F_i(q^2)$.
We shall discuss here spectral representations for the invariant amplitudes $F_i(q^2)$ and omit throughout the paper 
the subscript $i$.

As follows from the general properties of the time-ordered product, the function $F(q^2)$ is an analytic function in 
the complex $q^2$-plane
with the cut along the real axis \cite{barton,vol4} from a threshold $s_{\rm th}$ to $+\infty$ and satisfies the spectral
representation with an appropriate number $n$ of subtractions:
\begin{eqnarray}
\label{disp}
F(q^2)&=&F(0)+F'(0)q^2\nonumber\\
&&+\ldots+\frac{(q^2)^n}{\pi}\,\int\limits_{s_{\rm th}}^\infty \frac{ds}{s^n(s-q^2-i0)}\rho(s),
\end{eqnarray}
where $\rho(s)={\rm Im }~F(s+i 0)$. The subtractions are performed in order to provide the convergence of the
spectral integral or to satisfy constraints imposed by symmetries of the theory.

In QCD, one makes use of several expansions of the two-point functions. Important examples of such expansions are listed below:
(i) perturbative expansion in powers of $\alpha_s$ of the elastic correlation function 
\cite{voloshin,steinhauser_picc1,steinhauser_picc2,maierhofer,melnikov,brf};
(ii) rearrangement of the perturbation theory for the heavy-light correlation functions \cite{chetyrkin1,chetyrkin2}
via a replacement of the pole mass by the running mass of the heavy quark \cite{mass_tarrach,mass_chetyrkin,mass_melnikov}
to gain better convergence of the perturbation series;
(iii) expansion of the heavy-light correlation functions in the light-quark mass \cite{jamin,lms}.
In all these cases, higher-order spectral densities exhibit the appearance of nonintegrable divergences at the threshold.

Ref.~\cite{brf} gave a detailed analysis of the dispersion representations in the presence of threshold singularities, 
with the emphasis on vacuum polarization in QED: The authors made use of the known analytic results for 
the elastic QED two-point function (including the contributions of the continuum and the positronium poles) and its expansion in 
perturbation theory, in order to demonstrate their equivalence in the framework of dispersion relations. 

The novel feature of this paper is the procedure for handling nonintegrable threshold singularities in a more general 
case when only the discontinuity of the analytic function on the cut, but not the function itself, is known. 
We formulate an algorithm for solving the integral equation for the analytic function which arises when 
its discontinuity contains nonintegrable threshold singularities.  
Our analysis covers all three cases (i)-(iii) mentioned above and may be easily generalized to a nonintegrable threshold 
singularity of any form.  

\section{Threshold singularities in perturbation theory}
As an example consider radiative QED corrections to vacuum polarization in the vicinity of the $e^+e^-$ threshold.  
The continuum contribution to the spectral density of the vacuum polatization is known exactly and is given by 
the so-called Sommerfeld factor \cite{nosvvz}: 
\begin{eqnarray}
\label{rhocontinuum}
\rho(s)=\frac{\pi\alpha}{2}\frac{1}{1-\exp(-\pi\alpha/v)},
\quad v=\sqrt{1-\frac{4m^2}{s}}, 
\end{eqnarray}
$m$ the electron mass and $\alpha$ the electromagnetic coupling. 
Expanding this factor in powers of $\alpha$ yields
\begin{eqnarray}
\label{series}
\rho(s) &=& \frac{v}{2} + \frac{\pi\alpha}{4} +
\frac{1}{2} \sum\limits_{n=2,4\ldots}\frac{B_n}{n!}  \left( \pi \alpha \right)^n \frac{1}{v^{n-1}}, 
\end{eqnarray}
where $B_n$ are Bernoulli numbers ($B_2 = 1 /6$, $B_4 = -1 / 30$, ...).
The sum in Eq.~(\ref{series}) runs over even powers of $\alpha$: for $n=2$ it has
an integrable $1/v$ divergence at the threshold $v=0$, but starting with $n=4$, these singularities are nonintegrable and one should
specify the precise way to handle them.\footnote{In addition to the continuum contribution, there is also the contribution of 
positronium poles located below the threshold (see e.g. \cite{nosvvz}). 
Expanding the latter in powers of $\alpha$ and adding to the perturbative expansion 
of the continuum contribution, one gets the full perturbative spectral density. For our interest in this paper, only the perturbative expansion 
of the continuum is relevant: The contributions of the positronium poles are localized slightly below the threshold and fall on  
the threshold being expanded in powers of $\alpha$. Such contributions are integrable at the threshold and therefore are 
beyond the scope of our interest. To avoid confusion, we only recall that the positronium poles lead to nonzero 
small-circle contributions \cite{brf} [see Eq.~(\ref{B4})].} 

\begin{figure}[b!]
\begin{center}
\includegraphics[width=6.5cm]{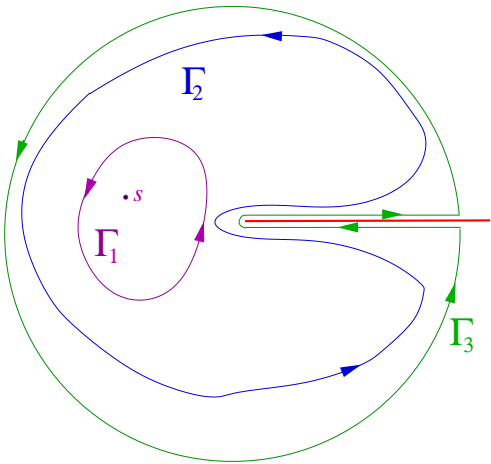}
\caption{\label{contour}
Deformations of the integration contour $\Gamma_1\to\Gamma_2\to\Gamma_3$ in Eq.~(\ref{cauchy}).
According to the Cauchy theorem, such deformations of the contour do not change the finite value of the contour integral.}
\end{center}
\end{figure}

To understand the proper way to proceed, we recall the following property of any analytic function:
{\it A contour integral of an analytic function over the region where the function is free of singularities is finite
and does not depend on the specific choice of the integration contour}. Let us look at the problem of the threshold
divergences from this perspective. We start with the Cauchy theorem
\begin{eqnarray}
\label{cauchy}
F(s)=F(0)+\frac{s}{2\pi i}\oint\limits_\Gamma
\frac{F(s')}{s'(s'-s)}ds'.
\end{eqnarray}
Here $\Gamma$ is any contour surrounding the point $s$ and located in the region where the function $F(s)$ is analytic.
We start with the contour $\Gamma_1$ (see Fig.~\ref{contour}).
Obviously, the integral is finite. We now start to deform the contour $\Gamma_1\to\Gamma_2\to \Gamma_3$ in the region of
analyticity of the function $F(s)$. Such deformations do not change the (finite) value of the integral.
Finally, we end up with the contour $\Gamma_3$ which embraces the cut from $s=4m^2$ to $s=\infty$, and the large circle. 
With the appropriate
number of subtractions, the large-circle integral vanishes, and we end up with the contour integral which embraces the cut. 
This
integral needs some care. It may be split into two parts:
(I) the integral over a small circle of radius $\epsilon$ around the threshold $s=4m^2$, and (II) the integral over the ``cut'' starting from
$s=4m^2+\epsilon$ to $\infty$ along the real $s$-axis. The sum of these two integrals is finite due to the general property of the
analytic functions mentioned above.
If the behaviour of the function $F(s)$ near the threshold is such that the integral of ${\rm Im}\, F(s)$ over the cut is finite,
then the small-circle integral vanished as $\epsilon\to 0$.
However, the situation changes for the case when ${\rm Im}\, F(s)$ has a nonintegrable singularity at the threshold: then,
the small-circle integral (I) also diverges for $\epsilon\to 0$,
making the sum of (I) and (II) finite for $\epsilon\to 0$.


We shall demonstrate this property for the general case of the singular threshold behaviour of the spectral density of the form 
${\rm Im}\, F(s+i 0)=\rho(s)\simeq 1/v^n$. According to the behaviour of the imaginary part of $F(s)$,
${\rm Im}\, F(s)=\rho(s) \to 1$ at large $s$, one subtraction in the spectral representation
for $F(s)$ is sufficient to guarantee that the large-contour integral vanishes.

For convenience, we introduce a dimensionless variable $z=s/4m^2$, such that the cut in $F(z)$
is located along the real $z$-axis from 1 to $\infty$, and write for the function $F(z)$ the following dispersion
representation with one subtraction (see also \cite{brf}):
\begin{eqnarray}
\label{B4}
F(z)&=&F(0)+
\frac{z}{2\pi i}\oint\limits_{R_\epsilon}\frac{F(z')}{z'(z'-z)}dz'\nonumber\\
&+&
\frac{z}{\pi}\int\limits_{1+\epsilon}^\infty
\frac{\rho(z')}{z'(z'-z-i0) } dz'.
\end{eqnarray}
where $R_\epsilon$ is the (clockwise) circle with radius $\epsilon$ and the center at $z=1$.
Equation (\ref{B4}) is an inhomogeneous integral equation for $F(z)$ and its solution is not fully trivial.

Technically, the cases of an even $n$ and an odd $n$ are slightly different, so we shall consider them separately. 

\subsection{Odd powers of $1/v$}
The spectral density of the elastic two-point function, Eq.~(\ref{series}), contains only odd inverse powers of $v$. 
Let us start with the case of $O(\alpha_s^4)$ correction, namely
\begin{eqnarray}
\rho_4(s) \propto \frac{1}{v^3}, \qquad v=\sqrt{1-\frac{4m^2}{s}}.
\label{rho4}
\end{eqnarray}
A detailed discussion of the above correction in the case of the hadronic vacuum polarization contribution to 
the electron ($g - 2$) can be found in Ref.~\cite{brf}.

We first analyse the contribution to $F_{3/2}(z)$ from the dispersion integral over the cut and thus see which parts of this contribution
are to be inserted in the small-circle integral. By a manipulation with the integrand,
\begin{eqnarray}
\label{2.6}
\frac{1}{1-\frac{z}{z'}} &=& \left(\frac{1}{1-\frac{z}{z'}}-\frac{1}{1-z}\right) +\frac{1}{1-z} \nonumber\\
& =& - \frac{z}{1-z} \frac{1 - \frac{1}{z'}}{1 - \frac{z}{z'}} + \frac{1}{1 - z},
\end{eqnarray}
we isolate the term singular in $\epsilon$ and obtain
\begin{eqnarray}
\label{F3}
&&F_{3/2}(z,\epsilon)\equiv
\frac{z}{\pi}\int\limits_{1+\epsilon}^\infty
\frac{1}{(1-1/z')^{3/2}z'(z'-z) } dz'     \nonumber\\
&&=
\frac{2z}{\pi(1-z)}\frac{1}{\sqrt{\epsilon}}
-\frac{2z}{\pi(1-z)}
-\frac{z}{1-z}F_{1/2}(z,0)\nonumber\\
&&\quad +O(\sqrt{\epsilon}),
\end{eqnarray}
where
\begin{eqnarray}
\label{F1}
F_{1/2}(z,\epsilon)&\equiv&
\frac{z}{\pi}\int\limits_{1+\epsilon}^\infty
\frac{1}{\left(1-1/z'\right)^{1/2}z'(z'-z-i 0) } dz', \nonumber\\
F_{1/2}(z,0)
&=& -\frac{2}{\pi}\sqrt{\frac{z}{z-1}}\log\left(\sqrt{-z}+\sqrt{1-z}\right).\nonumber\\
\end{eqnarray}
Since $F_{1/2}(z,\epsilon)$ is a regular function at $\epsilon\to 0$, we could replace in (\ref{F3}) 
$F_{1/2}(z,\epsilon)$ with $F_{1/2}(z,0)$. 

It is convenient to write (\ref{F3}) in the following form
\begin{eqnarray}
\label{F3expansion}
F_{3/2}(z,\epsilon) 
&=& \frac{2}{\pi} \frac{z}{1-z}\frac{1}{\sqrt{\epsilon}} \nonumber\\
&+&\frac{2}{\pi} \frac{z}{z-1}
\left[ 1-\sqrt{\frac{z}{z-1}}\mbox{log}\left( \sqrt{z}+\sqrt{z-1}\right) \right] \nonumber\\ 
&-& \left( \frac{z}{1-z} \right)^{3/2}+O(\sqrt{\epsilon}), 
\end{eqnarray}
where the second term on the r.h.s. is finite at the threshold. 

Let us write down again Eq.~(\ref{B4}):
\begin{eqnarray}
\label{Fa0}
F_{3/2}(z)=F_{3/2}(0)+\frac{z}{2\pi i}\oint\limits_{R_\epsilon}\frac{F_{3/2}(z')}{z'(z'-z)}dz'+F_{3/2}(z,\epsilon). \nonumber\\ 
\end{eqnarray}
Because of the Cauchy theorem, the function $F_{3/2}(z)$ does not depend on $\epsilon$. This means that the term $\sim\epsilon^{-1/2}$
generated by $F_{3/2}(z,\epsilon)$ cannot be a part of $F_{3/2}(z)$ and should cancel against the small-circle integral.
We shall now demonstrate that the cancellation of this divergent term is the only effect of the small-circle integral
and that it does not yield any contribution that remains finite in the limit $\epsilon\to 0$. In other words, 
{\it to obtain $F_{3/2}(z)$ one needs to
subtract from $F_{3/2}(z,\epsilon)$ all singular terms in $\epsilon$ and then send $\epsilon\to 0$}. 
To prove this statement we just
show that $F_{3/2}(z)$ obtained in this way satisfies Eq.~(\ref{Fa0}).

We turn to Eq.~(\ref{F3expansion}) and omit the first term in its r.h.s., which is singular in $\epsilon$. 
Then we check which of the remaining structures may give a nonvanishing contribution when integrated over the small circle.
The second term in the r.h.s.~of Eq.~(\ref{F3expansion}) is nonsingular at the threshold, 
as it can be easily checked by an expansion around $z = 1$.
Therefore, its contribution to the small-circle integral vanishes after the limit $\epsilon \to 0$ is taken.

Only the last term in the r.h.s.~of Eq.~(\ref{F3expansion}), which is singular at the threshold, 
can contribute to the small-circle integral. It is convenient to set $z'=1-\epsilon \, e^{i\phi}$ 
and take the $\phi$-integral from $\pi$ to $-\pi$ that corresponds to the clockwise contour integration
(the angle $\phi$ is measured from the negative direction of the $z$-axis; 
at $\phi=-\pi$ we are on the upper boundary of the cut, and at $\phi=\pi$ we are on its lower boundary), 
one gets
\begin{eqnarray}
I_{3/2}(z,\epsilon)&\equiv &-\frac{z}{2\pi i}\oint\limits_{R_\epsilon}\frac{\sqrt{z'}}{(1-z')^{3/2}}\frac{dz'}{z'-z}\nonumber\\ 
&&=\frac{z}{2\pi(1-z)\sqrt{\epsilon}}
\int\limits_{-\pi}^{\pi}d\phi ~ e^{-i \phi/2} + O(\sqrt{\epsilon})\nonumber\\ 
&&=-\frac{2}{\pi} \frac{z}{1-z} \frac{1}{\sqrt{\epsilon}} + O(\sqrt{\epsilon}),
\end{eqnarray}
where we made use of the integral $\int\limits_{-\pi}^{\pi} d\phi ~ e^{-i \phi/2} = 4$.
Thus, the small-circle integral $I_{3/2}(z,\epsilon)$ precisely cancels the singular contribution in $1/\sqrt{\epsilon}$ coming 
from $F_{3/2}(z,\epsilon)$ and does not develop any finite terms in the limit $\epsilon\to 0$.

Finally, in the case of the perturbative term (\ref{rho4}) we come to the following solution for $F_{3/2}(z)$:
\begin{eqnarray}
\label{Fa}
F_{3/2}(z)&=&F_{3/2}(0)+\frac{2}{\pi} \frac{z}{z-1}\nonumber\\
&&-\frac{2}{\pi} \frac{z}{z-1} \sqrt{\frac{z}{z-1}}\log(\sqrt{-z}+\sqrt{1-z}).\nonumber\\
\end{eqnarray}
It is straightforward to check that on the upper boundary of the cut along the real axis from $z=1$ to $+\infty$
[to get there one needs to set $z\to z+i 0$ for $z>1$], the imaginary part of $F(z)$ is indeed equal to $(1-1/z)^{-3/2}$.
The behaviour of $F_{3/2}(z)$ near the threshold $z=1$ reads  
\begin{eqnarray}
\label{Faexpansion}
&&F_{3/2}(z+i 0)-F_{3/2}(0)\nonumber\\ 
&&=i \left(\frac{z}{z-1}\right)^{3/2}-\frac{2}{3\pi}-\frac{2}{5\pi}(z-1)+O((z-1)^2),\nonumber\\
\end{eqnarray}
and $F_{3/2}(z)$ verifies Eq.~(\ref{B4}) in the limit $\epsilon\to 0$.

For higher-order contributions, the small-circle integrals produce singular terms containing a series of
inverse powers of $\sqrt{\epsilon}$ related to poles of the increasing order at the threshold $z=1$.

A recursive expression for $F_{n+\frac12}(z)$ for an arbitrary integer $n\ge 1$ reads (a derivation of this expression 
and a proof by induction of the cancellation 
of all $\epsilon$-dependent terms is given in Appendix \ref{AppA}):
\begin{eqnarray}
F_{n+\frac12}(z)-F_{n+\frac12}(0)
&=& \frac{z}{z-1}\left\{ F_{n-\frac12}(z)-F_{n-\frac12}(0)\frac{}{}\right\}\nonumber\\
&&+\frac{1}{\pi}\frac{z}{z-1}
\frac{1}{(n-\frac12)}. 
\end{eqnarray}


\subsection{Even powers of $1/v$}
In several practical applications, e.g., when considering the expansion of the heavy-light two-point correlation functions, 
one encounters spectral densities with threshold singularities of the type $1/v^{2n}$, with integer $n$ \cite{jamin}.\footnote{One should not confuse 
even powers of $1/v$ in the spectral density of the two-point function considered here and even powers of $1/v$ in the two-point function 
itself \cite{brf}: the latter emerge in the elastic two-point function in QED and correspond to the $\delta$-function and its derivatives in the 
spectral density and therefore do not represent nonintegrable threshold singularities.} 
Conceptually this case has no difference compared to the odd powers considered previously, but technically there 
are some subtleties.

The dispersion integral over the cut reads  
\begin{eqnarray}
F_n(z,\epsilon) &=& \frac{z}{\pi} \int_{1 + \epsilon}^\infty dz' \frac{1}{(1- \frac{1}{z'})^n} \frac{1}{z'(z'-z)}.
\end{eqnarray}
Using Eq.~(\ref{2.6}), one can obtain the following relation
\begin{eqnarray}
F_n(z,\epsilon) &=& \frac{z}{z-1} F_{n-1}(z,\epsilon) \nonumber\\
&&- \frac{1}{\pi} \frac{z}{z-1} 
\int_{1 + \epsilon}^\infty dz' \frac{1}{(1- \frac{1}{z'})^n} \frac{1}{z'^2} ~ , ~
\end{eqnarray}
that gives, after taking the integral, 
\begin{eqnarray}
\label{fnzepsrecursive}
F_n(z,\epsilon)  &=& \frac{z}{z-1} F_{n-1}(z,\epsilon) \nonumber\\
&&- 
\frac{1}{\pi} \frac{z}{z-1} \frac{  \left[ (1 + \frac{1}{\epsilon})^n - 1\right] \epsilon - 1}{(n-1)(1 + \epsilon)}. 
\end{eqnarray}
For $n>1$, the last term contains only positive powers of $1 / \epsilon$ (in particular, no constant term). 
As already explained above, the $\epsilon$-dependent piece of $F_n(z,\epsilon)$ cannot be a part of the analytic function $F_n(z)$. 
So, (\ref{fnzepsrecursive}) suggests that the analytic function $F_n(z)$ satisfies the relation 
\begin{eqnarray}
\label{rec}
F_n(z)-F_n(0)= \frac{z}{z-1} \left\{F_{n-1}(z) - F_{n-1}(0) \frac{}{}\right\}.
\end{eqnarray} 
For $n=1$, one finds 
\begin{eqnarray}
F_1(z,\epsilon)= 
-\frac{1}{\pi} \frac{z}{z-1}\log(1-z)
+\frac{1}{\pi} \frac{z}{z-1} \log(\epsilon)+O(\epsilon)\nonumber\\ 
\end{eqnarray}
and its $\epsilon$-independent part suggests 
\begin{eqnarray}
\label{f1}
F_1(z)-F_1(0)= 
-\frac{1}{\pi} \frac{z}{z-1}\log(1-z).  
\end{eqnarray}
Starting with this function, Eq.~(\ref{rec}) yields 
\begin{eqnarray}
\label{fn}
F_n(z)-F_n(0)=   -  \frac{1}{\pi} \left( \frac{z}{z - 1} \right)^n \mbox{log}(1-z).  
\end{eqnarray}
This solution is still our conjecture and we need to check that $F_n(z)$ 
given by (\ref{fn}) satisfies the integral equation (\ref{2.6}). 
To this end, we insert (\ref{fn}) in the small-circle integral and check that the only role of the small-circle integral is indeed 
to precisely cancel the singular $\epsilon$-dependent terms in $F_n(z,\epsilon)$. 

Let us illustrate how things work, e.g., for $n=2$:  
\begin{eqnarray}
F_2(z,\epsilon) & = &  - \frac{1}{\pi} \left( \frac{z}{z - 1} \right)^2 \log(1-z) \nonumber \\
&& +\frac{1}{\pi} \left( \frac{z}{z - 1} \right)^2 \log(\epsilon)
+ \frac{1}{\pi} \frac{z}{1 - z} \frac{1}{\epsilon}+O(\epsilon).
\nonumber\\
\end{eqnarray}
We now have to insert its $\epsilon$-independent part, which as expected gives the solution $F_2(z)$, in the contour integral.
Substituting $z'=1-\epsilon \exp(i \phi)$, such that the clockwise $\phi$ integral runs from $\pi$ to $-\pi$, we obtain 
\begin{eqnarray}
I_2(z,\epsilon)&=&-\frac{z}{1 - z}\nonumber\\ 
&&\times\frac{1}{2\pi i} 
\int_{-\pi}^{\pi} \frac{d\phi}{\pi}
\frac{\epsilon i e^{i \phi}}{\epsilon^2 e^{2i \phi}} 
\frac{1 - \epsilon e^{i \phi}}{1 - \epsilon e^{i \phi} \frac{1}{1 - z}}
\left[ \mbox{log}(\epsilon) + i \phi) \right]\nonumber\\
\end{eqnarray}
yielding
\begin{eqnarray}
I_2(z,\epsilon)=- \frac{1}{\pi} \left( \frac{z}{z - 1} \right)^2 \log(\epsilon) - \frac{1}{\pi} \frac{z}{1 - z} \frac{1}{\epsilon}+O(\epsilon).\nonumber\\
\end{eqnarray}
The small-circle integral cancels all $\epsilon$-singular terms in the integral over the cut and 
does not add any finite contributions for $\epsilon\to 0$. 

A proof by induction of the cancellation of the $\epsilon$-dependent terms 
(not only singular but all $\epsilon$-dependent terms) in the sum $F_n(z,\epsilon)+I_n(z,\epsilon)$ 
for any integer $n$ is given in Appendix \ref{AppB}. So, our procedure indeed yields the solution 
of the integral equation (\ref{B4}) in the form (\ref{fn}). 
%

Before closing this Section, we emphasize that it is not necessary to calculate the small-circle integrals explicitly; 
it is sufficient to calculate the integral over the cut in Eq.~(\ref{B4}): 
The $\epsilon$-independent part gives the analytic function $F(z)$, while (due to the Cauchy theorem) 
the $\epsilon$-dependent terms in the dispersion integral in Eq.~(\ref{B4}) are precisely cancelled by  
the small-circle contribution. As we have seen, the cancellation of the $\epsilon$-dependent terms is the only role of 
the small-circle integral in the case of nonintegrable threshold singularities; 
it does not provide any finite contributions in the limit $\epsilon\to 0$.  


\section{Conclusions}
We discussed the way to handle properly the threshold divergences arising in perturbation theory for two-point functions.
Our results are as follows:

\vspace{.1cm}
\begin{itemize}
\item[(i)]
Taking a proper account of the small-circle integral around the threshold leads to the ``surface term'' that exactly cancels 
the threshold divergence of the spectral integral. Thus, a properly defined dispersion representation for $F(s)$ 
(and in fact for any analytic function) at each order of perturbative expansion is finite and does not have any threshold divergence. 

It should be understood that when the small-circle integral provides a nonzero contribution, the Cauchy theorem 
leads to the integral equation for the function $F(s)$ even if its discontinuity $\rho(s)$ on the cut is known analytically. 
We formulated an algorithm for solving this integral equation: namely, one should calculate the integral over the cut and isolate
in this integral the $\epsilon$-singular and the $\epsilon$-regular parts; 
the $\epsilon$-regular part gives the desired solution of the integral equation, while the 
$\epsilon$-singular part precisely cancels the small-circle integral. Noteworthy, the only role of the small-circle integral 
is the cancellation of the $\epsilon$-singularities. At each order of the perturbative expansion, the small-circle integral does 
not lead to finite contributions. 

We presented the analysis of nonintegrable threshold 
divergences of the general structure $\rho(s)\sim 1/v^n$, where $n$ can be both even and odd integer numbers $n \ge 2$. 
Moreover, it should be clear that the proposed algorithm is applicable to any singular threshold behaviour of the 
spectral density. 

Let us emphasize that thanks to the Cauchy theorem the explicit calculation of the small-circle integrals is not required. 
It is sufficient to isolate all terms singular in $\epsilon$ in the dispersion integral of $\rho(s)$: 
the small-circle integral is precisely equal to the singular part of the integral over the cut taken with an opposite sign. 
Noteworthy, the small-circle contribution does not yield any finite remnants in the limit $\epsilon\to 0$.

\item[(ii)] 
We would like to mention once more a detailed analysis of the threshold singularities in the dispersion representation for 
the vacuum polarization in QED \cite{brf}, in which case the spectral function contains contributions of the continuum states and 
the bound positronium states. The analysis of \cite{brf} made use of the knowledge of the analytic results of 
both the vacuum polarization and its imaginary part to demonstrate the relevance of the small-circle integrals in the 
dispersion representation for the vacuum polarization. Notice, that the finite small-circle contributions 
discussed in \cite{brf} come from the positronium poles. The latter are localized near the threshold and their 
contributions to the vacuum polarization are integrable. 

The goal of our analysis is the algorithm of treating nonintegrable threshold singularities that emerge in perturbation theory; 
our algorithm allows one to obtain the analytic function as the solution of the integral equation and does not require the 
a priori knowledge of this analytic function.  
\end{itemize}

In conclusion, 
in higher orders of the perturbation theory, nonintegrable threshold singularities given by powers of $1/v$ emerge. 
Our algorithm allows one to handle properly nonintegrable threshold singularities of two-point functions 
and leads to finite perturbative terms at each order; therefore, it may have a broad applicability. 
In particular, our algorithm opens the possibility of applying perturbative series to the investigation of 
ground-state hadrons within the machinery of QCD sum rules.

\vspace{.2cm}
\noindent{\it{\bf Acknowledgements.}} 
We are grateful to M.~Steinhauser for illuminating comments concerning the application of his program $rhad$
\cite{steinhauser_picc2}. D.~M. was supported by RFBR under project 19-52-15022.
S.~S. thanks the Italian Ministry of Research (MIUR) for support under the grant PRIN 20172LNEEZ.

\appendix
\section{
\label{AppA}Proof by induction for $F_{n+\frac12}$ [odd powers $1/v^{2n+1}$]}

We start with the integral over the cut 
\begin{eqnarray}
\label{bfnzeps}
F_{n+\frac12}(z,\epsilon)= \frac{z}{\pi}\int_{1 + \epsilon}^\infty dz' \frac{1}{(1- \frac{1}{z'})^{n+\frac12}} \frac{1}{z'(z'-z)}. 
\end{eqnarray}
By virtue of (\ref{2.6}), we come to a recursive relation 
\begin{eqnarray}
\label{bfn}
F_{n+\frac12}(z,\epsilon)= \frac{z}{z-1} F_{n-\frac12}(z,\epsilon)
+ \frac{1}{\pi}\frac{z}{1 - z}\Delta F_{n+\frac12}(\epsilon),\nonumber\\
\end{eqnarray}
where 
\begin{eqnarray}
\Delta F_{n+\frac12}(\epsilon)&=&
\int_{1 + \epsilon}^\infty dz' \frac{1}{(1- \frac{1}{z'})^{n+\frac12}} \frac{1}{z'^2}\nonumber\\
&=&
\frac{\left(\frac{\epsilon}{1+\epsilon}\right)^{\frac12-n}-1}{n-\frac12}.
\end{eqnarray}
This function is represented by an infinite sum of powers of $\sqrt{\epsilon}$, 
the most singular term being of the order ${\epsilon}^{-(n-\frac12)}$. 
Noteworthy, the $\epsilon$-independent term in $\Delta F_{n+\frac12}(\epsilon)$ is 
\begin{eqnarray}
-\frac{1}{(n-\frac12)}.
\end{eqnarray} 
We now conjecture that the $\epsilon$-independent part of the recursive relation (\ref{bfn}) yields the 
recursive relation for $F_{n+\frac12}(z)$: 
\begin{eqnarray}
\label{bfn2}
F_{n+\frac12}(z)-F_{n+\frac12}(0)
&=& \frac{z}{z-1}\left\{ F_{n-\frac12}(z)-F_{n-\frac12}(0)\frac{}{}\right\}\nonumber\\
&&+\frac{1}{\pi}\frac{z}{z-1}
\frac{1}{(n-\frac12)},\quad n\ge 1,\nonumber\\
\end{eqnarray}
where the knowledge of the $\epsilon$-independent term in $\Delta F_{n+\frac12}(\epsilon)$ is used.  

The analytic solution $F_{\frac12}(z)$ is easily found:  
\begin{eqnarray}
F_{\frac12}(z)=F_{\frac12}(0) -\frac{2}{\pi}\sqrt{\frac{z}{z-1}}\log\left(\sqrt{-z}+\sqrt{1-z}\right).
\end{eqnarray}
Solving (\ref{bfn2}) iteratively, we find the conjectured solution $F_{n+\frac12}(z)$. 
We still have to prove that this function indeed satisfies the integral equation (\ref{B4}). 
To this end, we need to calculate the small-circle integral of the function $F_{n+\frac12}(z)$, see Eq.~(\ref{B4}). 

Using (\ref{2.6}), the small-circle integral satisfies a recursive relation similar to (\ref{bfn}): 
\begin{eqnarray}
I_{n+\frac12}(z,\epsilon) &=& \frac{z}{z-1} I_{n-\frac12}(z,\epsilon)+ \frac{1}{\pi}\frac{z}{1 - z}\Delta I_{n+\frac12}(\epsilon),\nonumber\\
\end{eqnarray}
where 
\begin{eqnarray}
\label{A7}
\Delta I_{n+\frac12}(\epsilon)&=&
\frac{1}{2\pi i}\oint_{1 + \epsilon} dz' F_{n+\frac12}(z')\frac{1}{z'^2}. 
\end{eqnarray}
To calculate this small-circle integral, we expand the integrand near $z=1$. Making use of the expansion 
\begin{eqnarray}
F_{n+\frac12}(z)&=&(-1)^n\left(\frac{z}{1-z}\right)^{n+\frac12}+O((z-1)^0)
\nonumber\\
\end{eqnarray}
and expanding also $1/z^2$ near $z=1$, we find 
\begin{eqnarray}
\Delta F_{n+\frac12}(\epsilon)+\Delta I_{n+\frac12}(\epsilon)=-\frac{1}{(n-\frac12)}, 
\end{eqnarray}
yielding  
\begin{eqnarray}
&&F_{n+\frac12}(z,\epsilon)+I_{n+\frac12}(z,\epsilon)\nonumber\\
&&\quad=\frac{z}{z-1}\left\{ F_{n-\frac12}(z,\epsilon)+I_{n-\frac12}(z,\epsilon) \right\}
\nonumber\\
&&\qquad
+\frac{z}{\pi(z-1)}\frac{1}{(n-\frac12)}.
\end{eqnarray}
This equation coincides with Eq.~(\ref{bfn2}) for 
\begin{eqnarray}
F_{n+\frac12}(z)-F_{n+\frac12}(0)
=F_{n+\frac12}(z,\epsilon)+I_{n+\frac12}(z,\epsilon)
\end{eqnarray}
and verifies by induction that the sum $F_{n+\frac12}(z,\epsilon)+I_{n+\frac12}(z,\epsilon)$ does not depend on $\epsilon$
(since $F_{\frac12}(z)-F_{\frac12}(0)$ is $\epsilon$-independent). 
We therefore prove our conjecture that the solution $F_{n+\frac12}(z)$ is obtained from the dispersion integral 
over the cut, $F_{n+\frac12}(z,\epsilon)$, by omitting {\it all} $\epsilon$-dependent terms. 
Let us emphasize that the {\it only} role of the small-circle 
integral $I_{n+\frac12}(z,\epsilon)$ is to cancel the $\epsilon$-dependent terms 
in $F_{n+\frac12}(z,\epsilon)$; it leaves no other traces.

\section{\label{AppB}
Proof by induction for $F_n$ [even powers $1/v^{2n}$]}
This appendix gives the analysis for the case $F_{n}$ which has some subtleties 
compared to the analysis for $F_{n+\frac12}$ of Appendix \ref{AppA}.

We start with the integral over the cut 
\begin{eqnarray}
\label{afnzeps}
F_n(z,\epsilon)= \frac{z}{\pi}\int_{1 + \epsilon}^\infty dz' \frac{1}{(1- \frac{1}{z'})^n} \frac{1}{z'(z'-z)}. 
\end{eqnarray}
By virtue of (\ref{2.6}), one obtains a recursive relation 
\begin{eqnarray}
\label{afn}
F_n(z,\epsilon)= \frac{z}{z-1} F_{n-1}(z,\epsilon)+ \frac{1}{\pi}\frac{z}{1 - z}\Delta F_n(\epsilon),
\end{eqnarray}
where 
\begin{eqnarray}
\Delta F_n(\epsilon)&=&
\int_{1 + \epsilon}^\infty dz' \frac{1}{(1- \frac{1}{z'})^n} \frac{1}{z'^2}\nonumber\\
&=&
\frac{  \left[ (1 + \frac{1}{\epsilon})^n - 1\right] \epsilon - 1}{(n-1)(1 + \epsilon)}.
\end{eqnarray}
For integer $n$, this function is represented by a finite sum of poles in $1/\epsilon$ of the increasing order; 
it contains no regular structures in $\epsilon$.   

Solving (\ref{afn}) for the $\epsilon$-independent part and making use of the known expression for $F_1(z)$, 
we obtained $F_n(z)$ in the form [see (\ref{fn})]:    
\begin{eqnarray}
\label{afnsolution}
F_n(z)=F_n(0)  - \frac{1}{\pi} \left( \frac{z}{z - 1} \right)^n \log(1-z). 
\end{eqnarray}
This is still our conjecture and we have to check that this function satisfies the integral equation (\ref{B4}).

Let us calculate the small-circle integral for $F_n(z)$ given by (\ref{afnsolution}). 
Using again (\ref{2.6}), we obtain also for the small-circle integral a recursive relation
\begin{eqnarray}
I_n(z,\epsilon) &=& \frac{z}{z-1} I_{n-1}(z,\epsilon)+ \frac{1}{\pi}\frac{z}{1 - z}\Delta I_n(\epsilon),\nonumber\\
\end{eqnarray}
where 
\begin{eqnarray}
\label{dIn}
\Delta I_n(\epsilon)&=&
\frac{1}{2\pi i}\oint_{1 + \epsilon} dz'\log(1-z') \frac{1}{(1- \frac{1}{z'})^n} \frac{1}{z'^2}\nonumber\\
&=&
-\frac{  \left[ (1 + \frac{1}{\epsilon})^n - 1\right] \epsilon - 1}{(n-1)(1 + \epsilon)}=-\Delta F_n(\epsilon), \nonumber\\
\end{eqnarray}
Thus, $F_n(z,\epsilon)+I_n(z,\epsilon)$ satisifes the recursive relation 
\begin{eqnarray}
\label{B7}
F_n(z,\epsilon)+I_n(z,\epsilon)=\frac{z}{z-1}\left\{F_{n-1}(z,\epsilon)+I_{n-1}(z,\epsilon) \frac{}{}\right\}.\nonumber\\
\end{eqnarray}
For $n=1$ we find
\begin{eqnarray}
F_1(z,\epsilon)&=& 
-\frac{1}{\pi} \frac{z}{z-1}\log(1-z)
+\frac{1}{\pi} \frac{z}{z-1} \log(\epsilon)\nonumber\\
&&+\frac{z}{(z-1)^2}\epsilon+\frac{z}{2(z-1)^3}\epsilon^2+O(\epsilon^3)
\end{eqnarray}
and 
\begin{eqnarray}
I_1(z,\epsilon)&=& 
-\frac{1}{\pi} \frac{z}{z-1} \log(\epsilon)\nonumber\\
&&-\frac{z}{(z-1)^2}\epsilon-\frac{z}{2(z-1)^3}\epsilon^2+O(\epsilon^3),
\end{eqnarray}
leading to 
\begin{eqnarray}
F_1(z,\epsilon)+I_1(z,\epsilon)=-\frac{1}{\pi}\frac{z}{z-1}\log(1-z).
\end{eqnarray}
Let us emphasize again that not only singular in $\epsilon$ terms, but 
also powers of $\epsilon$ cancel in the sum, so 
$F_1(z,\epsilon)+I_1(z,\epsilon)$ does not depend on $\epsilon$ at all.  
As follows from (\ref{B7}), for any $n$ the sum $F_n(z,\epsilon)+I_n(z,\epsilon)$ also does not depend on $\epsilon$ and has the form 
\begin{eqnarray}
F_n(z,\epsilon)+I_n(z,\epsilon)=-\frac{1}{\pi}\left(\frac{z}{z-1}\right)^n\log(1-z).
\end{eqnarray}
So our conjectured solution (\ref{afnsolution}) indeed solves the integral equation (\ref{B4}).


\end{document}